\documentclass[showpacs,aps]{revtex4}
\usepackage{graphicx}
\usepackage{dcolumn}
\usepackage{bm}

\begin{document}

%

\title{Excited electron contribution to the 
$\rm e^+e^-\rightarrow\gamma\gamma\;$  
cross-section}

\author{Brigitte Vachon}
 \email{Brigitte.Vachon@cern.ch}
 \affiliation{University of Victoria, 
              Department of Physics and Astronomy, 
              P.O. Box 3055, Victoria, B.C., V8W 3P6, Canada.}

\date{\today}

\begin{abstract}
The differential cross-section for the process 
$\rm e^+e^-\rightarrow\gamma\gamma\; $ is calculated assuming the presence
of excited electrons with a chiral magnetic coupling.  
This calculation permits constraining the excited electron
coupling using the same theoretical framework as the one generally used for 
direct production searches.
\end{abstract}


\maketitle

\begin{figure}[h]
  \begin{center}
    \includegraphics[width=8.5cm]{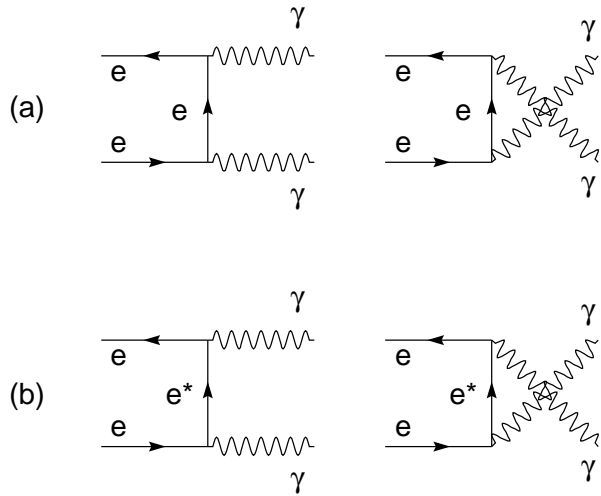}
    \caption{Diagrams showing the (a) Standard Model and (b) excited
      electron contributions to the process 
      $\rm e^+e^-\rightarrow\gamma\gamma$. }
    \label{feyn_indirect}
  \end{center}
\end{figure}

Excited electrons
could be produced directly in $\rm e^+e^-$ collisions.
They are also expected to alter the 
$\rm e^+e^-\rightarrow\gamma\gamma\;$ differential cross-section
via the diagrams shown in Fig.~\ref{feyn_indirect}.
Currently, the limits on the size of the 
$\rm e^*e\,\gamma$ coupling obtained from 
direct production~\cite{LEP_direct_search}
searches are generally calculated using one form 
of the $\rm e^*e\,\gamma$ interaction
whereas limits 
extracted from the $\rm e^+e^-\rightarrow\gamma\gamma\;$ differential 
cross-section~\cite{LEP_indirect_search} use a different form.
By using a common theoretical framework, the
$\rm e^+e^-\rightarrow\gamma\gamma\;$ process can be used to extend
the limits on the $\rm e^*e\,\gamma$ coupling strength for excited electron
masses beyond the kinematically allowed region of direct production.
This paper presents a brief overview of both 
forms of the $\rm e^*e\,\gamma$ interaction and the analytic 
expression for the $\rm e^+e^-\rightarrow\gamma\gamma\;$ differential 
cross-section derived in the framework used by direct production searches.
This differential cross-section is compared to the form currently in use
and the impact on the limits of the 
$\rm e^*e\,\gamma$ coupling strength is discussed.

Indirect searches, where the existence of $\rm e^*$ is 
inferred from deviations in the $\rm e^+~e^-~\rightarrow~\gamma~\gamma$ 
differential cross-section, usually express limits on the
$\rm e^*e\gamma$ coupling strength 
assuming a general extension of the Standard Model.  
The interaction 
between an excited lepton, a lepton and a gauge boson ($\ell^*\ell\,V$) 
is described by the simplest gauge invariant
form of the interaction Lagrangian~\cite{Litke}

\begin{equation}
\label{vertex_indirect} 
{\cal L}_{\ell^*\ell\,V} = \frac{e}{2}\;
\frac{\kappa}{M_{\ell^*}}\; \bar{\ell}^*\, \sigma_{\mu\nu}\, 
\ell\, \rm F^{\mu\nu} + h.c.
\end{equation}
where $\rm F^{\mu\nu}$ denotes the electromagnetic field tensor, 
$\sigma_{\mu\nu}$ is the covariant bilinear tensor and 
$M_{\ell^*}$ is the mass of the excited lepton.
The parameter $\kappa$
is a measure of the coupling strength.  
The $\rm e^+e^-\rightarrow\gamma\gamma$ differential
cross-section  
using this purely magnetic coupling is explicitly 
calculated in \cite{Litke} and given by 

\begin{eqnarray}
\label{dxs_litke}
  \left( \frac{d\sigma}{d\Omega} \right)_{\rm e^*} = & &
       \left( \frac{d\sigma}{d\Omega} \right)_{\rm QED} \nonumber \\
   & & +\; \alpha^2 \left\{  
       \frac{1}{2} \left( \frac{\kappa}{M_{\rm e^*}} \right)^4 
       (E^2\sin^2\theta + M_{\rm e^*}^2) 
       \left(
           \frac{q^4}{\left( q^2 - M_{\rm e^*}^2 \right)^2} + 
           \frac{q^{\,\prime 4}}{\left( q^{\,\prime 2} - M_{\rm e^*}^2\right)^2} 
       \right) \right.\nonumber \\
   & & +\; 4 \left( \frac{\kappa}{M_{\rm e^*}} \right)^4 
       \frac{M_{\rm e^*}^2E^4\sin^2\theta}{(q^2-M_{\rm e^*}^2)(q^{\,\prime 2}-M_{\rm e^*}^2)} \nonumber \\
   & & +\; \left. \left( \frac{\kappa}{M_{\rm e^*}} \right)^2
       \left[ \frac{q^2}{q^2-M_{\rm e^*}^2} + 
              \frac{q^{\,\prime 2}}{q^{\,\prime 2}-M_{\rm e^*}^2} +
              E^2\sin^2\theta\left( \frac{1}{q^2-M_{\rm e^*}^2} +
                                    \frac{1}{q^{\,\prime 2}-M_{\rm e^*}^2} \right)
       \right] \right\}
\end{eqnarray}
where $\left( \frac{d\sigma}{d\Omega} \right)_{\rm QED}$ is the
Born level Standard Model differential cross-section,  
$\theta$ is the polar angle of one of the photons with respect to the incoming
electron, $E$ is the beam energy ($E = \sqrt{s}/2$), 
$q^2 = -2E^2(1-\cos\theta)$ and $q^{\,\prime 2} = -2E^2(1+\cos\theta)$.  
Since the two 
outgoing photons are indistinguishable, $\cos\theta$ is defined to 
be positive. 
Limits on the strength of the $\rm e^*e\gamma$ coupling, 
$\kappa$, are expressed 
as a function of $M_{\rm e^*}$~\cite{LEP_indirect_search}.

The interaction Lagrangian of Equation~\ref{vertex_indirect} 
leads to
large contributions to the anomalous magnetic moment of electrons and 
muons and the coupling is therefore severely constrained by 
existing g-2 precision measurements~\cite{Renard,Brodsky}.
In fact, limits from g-2 measurements are approximately an order 
of magnitude better than limits from $\rm e^+e^-\rightarrow\gamma\gamma$ 
calculated using Equation~\ref{dxs_litke}.

Limits on the $\rm e^*e\gamma$ coupling strength from the search 
for directly produced excited leptons are 
calculated using a different theoretical framework.
The effective Lagrangian describing the $\ell^*\ell\,V$ interaction
is chosen to have a chiral symmetry which protects Standard
Model leptons from acquiring large anomalous magnetic moments.
This Lagrangian is generally written 
as~\cite{BDK,Cabibbo,Kuhn,Hagiwara,Boudjema}

\begin{equation} 
\label{eff_Lagrangian}
 {\cal L}_{\ell^*\ell\,V} = \frac{1}{2\Lambda}\: \bar{\ell}^* \sigma^{\mu\nu}
 \left[ g f\frac{\bm{\tau}}{2}\bm{W}_{\!\!\mu\nu} + g^{\,\prime} 
 f^{\,\prime} \frac{Y}{2} B_{\mu\nu} 
 \right] \ell_L + \rm h.c. 
\end{equation}
where $g$ and $g^{\,\prime}$ are the usual Standard Model couplings, 
the tensors $\bm{W}_{\!\!\mu\nu}$ and $B_{\mu\nu}$
represent the Standard Model
gauge-invariant field tensors,
$\bm{\tau}$ denotes 
the Pauli matrices and $Y$ is the weak hypercharge.
The compositeness scale is set by the parameter $\Lambda$ which
has units of energy.  The size of the $\ell^*\ell\,V$ coupling is 
governed by the constants $f$ and $f^{\,\prime}$ which can be
interpreted as weight factors associated to the different gauge groups.
In the physical basis, the Lagrangian of Equation \ref{eff_Lagrangian} 
leads to the following chiral magnetic vertex~\cite{BDK}

\begin{equation}
\label{vertex_direct}
 \Gamma_\mu^{\ell^*\!\ell\,V} = \frac{e}{2\Lambda}\:f_V\: q^\nu \sigma_{\mu\nu}
 (1-\gamma_5) 
\end{equation}
where $q^\nu$ is the momentum of the gauge boson and $f_V$  represents
a particular combination of the constants $f$ and $f^{\,\prime}$ for a 
given boson interaction.  For an excited electron
coupling to photon ($V=\gamma$)~\cite{BDK}, 
$f_\gamma = -\frac{1}{2}(f+f^{\,\prime})$.
In this model, experimental limits on the $\ell^*\ell\,V$ coupling 
are usually expressed in terms of limits on  
$|f|/\Lambda$ as a function of the excited lepton mass
for the hypothesis~\footnote{For the specific $\rm e^*e\gamma$
interaction the photon 
coupling $f_\gamma$ is zero for the hypothesis 
$f=-f^{\,\prime}$ but the coupling
to $Z^0$ and $W^{\pm}$ does not vanish~\cite{BDK}.} 
$f=f^{\,\prime}$ or $f=-f^{\,\prime}$\cite{LEP_direct_search}.

Using this chiral magnetic interaction, the coupling is 
less severely constrained since contributions to the 
electron and muon anomalous magnetic moment are suppressed.
It still however permits observable deviations in the process
$\rm e^+e^-\rightarrow\gamma\gamma$ which are not excluded by 
g-2 measurements.
In addition, limits from indirect searches expressed in this framework
could be easily compared and combined with limits coming from direct searches.
To achieve this, deviations from the Standard Model 
$\rm e^+e^-\rightarrow\gamma\gamma$ 
differential cross-section must be calculated assuming a
chiral magnetic $\rm e^*e\gamma$ coupling.

With the existence of an excited electron, the four diagrams shown 
in Fig.~\ref{feyn_indirect},
contribute to the Born level production of 
$\rm e^+e^-\rightarrow\gamma\gamma$ events.
The differential cross-section 
is calculated using the $\ell^*\ell\,V$ vertex given in 
Equation~\ref{vertex_direct} and combined with the standard QED 
interaction $e\,\bar{\ell}\, \gamma_\mu\, \ell \rm A_\mu$.  The 
excited electron propagator is taken to be the usual fermion
expression with a mass $M_{\rm e^*}$.  
Summing over the outgoing photon polarizations and neglecting the 
mass of the electron, the resulting differential cross-section is

\begin{eqnarray}
\label{dxs_mine}
  \left( \frac{d\sigma}{d\Omega} \right)_{\rm e^*} & = & 
  \left( \frac{d\sigma}{d\Omega} \right)_{\rm QED}  \nonumber \\
   & & +\;
  \frac{\alpha^2}{4}\frac{f_\gamma^4}{\Lambda^4}E^2\sin^2\theta
  \left[ \frac{q^4}{\left( q^2 - M_{\rm e^*}^2 \right)^2} + 
	 \frac{q^{\,\prime 4}}{\left( q^{\,\prime 2} - M_{\rm e^*}^2
  \right)^2}
  \right] \nonumber \\
   & & -\; 
   \frac{\alpha^2}{8E^2}\frac{f_\gamma^2}{\Lambda^2} 
   \left[ \frac{q^4}{\left( q^2 - M_{\rm e^*}^2 \right)} + 
	 \frac{q^{\,\prime 4}}{\left( q^{\,\prime 2} - M_{\rm e^*}^2
  \right)}
  \right]
\end{eqnarray}
where the same notation as for Equation~\ref{dxs_litke} is used.
Terms of order $\left( f_\gamma/\Lambda \right)^2$ in
Equation~\ref{dxs_mine} describe the interference between the excited
electron and Standard Model diagrams.

\begin{figure}
  \begin{center}
    \includegraphics[width=8.5cm]{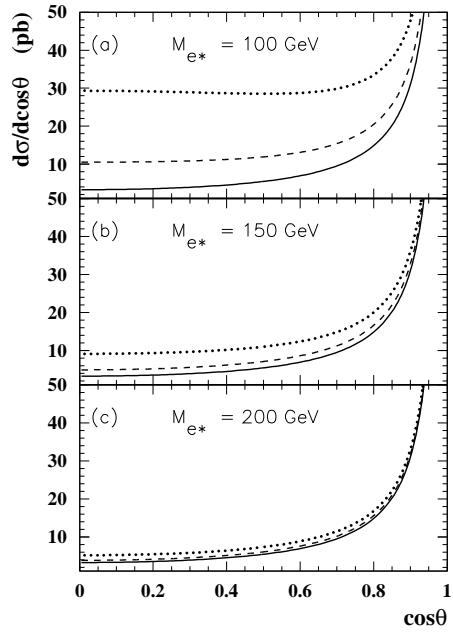}
    \caption{ Differential cross-section for the process 
      $\rm e^+e^-\rightarrow\gamma\gamma$. 
      The solid lines represent the Standard
      Model Born level differential cross-section. 
      The dotted lines show the prediction obtained assuming  
      the presence of an excited electron with purely magnetic
      coupling assuming $\kappa = 1$
      and the dashed lines represent the prediction from a chiral 
      magnetic coupling with $f_\gamma = 1$ and $\Lambda = M_{\rm e^*}$.
      The plot was generated using a center-of-mass energy of 
      200~GeV and with an excited electron mass of (a) 100~GeV,
      (b) 150~GeV and (c) 200~GeV. }
    \label{plot_dxs}
  \end{center}
\end{figure}

Figure~\ref{plot_dxs} shows the Standard Model Born level
differential cross-section for 
the process $\rm e^+e^-\rightarrow\gamma\gamma$ compared with 
the prediction obtained using a purely magnetic 
(Equation~\ref{dxs_litke})
and chiral magnetic (Equation~\ref{dxs_mine}) $\rm e^*e\gamma$ coupling.
For comparison purposes, the coupling parameters
are set to $\kappa = 1$ in Equation~\ref{dxs_litke}, 
and $f_\gamma = 1$ and $\Lambda = M_{\rm e^*}$ in Equation~\ref{dxs_mine}.
The differential cross-section assuming a purely magnetic coupling 
is larger than the one assuming a chiral magnetic coupling.

In summary, the differential cross-section for the process 
$\rm e^+e^-\rightarrow\gamma\gamma$ assuming the existence of
an excited electron with chiral magnetic coupling has been calculated.
The differential 
cross-section predictions from chiral magnetic and purely magnetic
couplings are compared.
Limits on the size of the $\rm e^*e\gamma$ coupling 
determined from the $\rm e^+e^-\rightarrow\gamma\gamma$ differential
cross-section can be combined 
with limits obtained from direct production searches when the chiral
magnetic form is used.

\bigskip
I would like to thank Robert McPherson, Randall Sobie and Matt Dobbs
for their useful comments and careful reading of the manuscript.
I am also grateful to Kirsten Sachs for stimulating discussions.
This work was supported by the Natural Sciences and Engineering 
Research Council of Canada.


%
%

\begin{references}
\bibitem{LEP_direct_search}
\setlength{\itemsep}{2.0mm}
   L3 Collaboration, M.~Acciarri {\it et al.}, 
   Phys.\ Lett.\ {\bf B} (to be published), CERN-EP/2000-143;

   OPAL Collaboration, G.~Abbiendi {\it et al.},
   Eur.\ Phys.\ J.\ C {\bf 14}, 73 (2000);

   DELPHI Collaboration, P.~Abreu {\it et al.},
   Phys.\ Lett.\ B {\bf 393}, 245 (1997);

   ALEPH Collaboration, D.~Buskulic {\it et al.},
   Phys.\ Lett.\ B {\bf 385}, 445 (1996).


\bibitem{LEP_indirect_search}
   DELPHI Collaboration, P.~Abreu {\it et al.},
   Phys.\ Lett.\ B {\bf 491}, 67 (2000);

   L3 Collaboration, M.~Acciarri {\it et al.},
   Phys.\ Lett.\ B {\bf 475}, 198 (2000);

   OPAL Collaboration, G.~Abbiendi {\it et al.},
   Phys.\ Lett.\ B {\bf 465}, 303 (1999);

   ALEPH Collaboration, R.~Barate {\it et al.},
   Phys.\ Lett.\ B {\bf 429}, 201 (1998);


\bibitem{Litke}
   A.~Litke, Ph.D. thesis, Harvard University, 1970.

\bibitem{Renard}
   F.~M.~Renard,
   Phys.\ Lett.\ B {\bf 116}, 264 (1982).

\bibitem{Brodsky}
   S.~J.~Brodsky and S.~D.~Drell,
   Phys.\ Rev.\ D {\bf 22}, 2236 (1980).


\bibitem{BDK}
   F.~Boudjema, A.~Djouadi, and J.~L.~Kneur,
   Z.\ Phys.\ C {\bf 57}, 425 (1993).

\bibitem{Cabibbo}
   N.~Cabibbo, L.~Maiani, and Y.~Srivastava,
   Phys.\ Lett.\ B {\bf 139}, 459 (1984).

\bibitem{Kuhn}
   J.~K\"{u}hn and P.~Zerwas,
   Phys.\ Lett.\ B {\bf 147}, 189 (1984).

\bibitem{Hagiwara}
   K.~Hagiwara, D.~Zeppenfeld, and S.~Komamiya,
   Z.\ Phys.\ C {\bf 29}, 115 (1985).

\bibitem{Boudjema}
   F.~Boudjema and A.~Djouadi,
   Phys.\ Lett.\ B {\bf 240}, 485 (1990).


\end{references}
\end{document}